\documentclass[11pt,twoside]{article}
\usepackage{asp2010}

\resetcounters

\begin{document}

\title{Runaway growth of fractal dust grains}
\author{Lars Mattsson$^1$, and Joakim D. Munkhammar$^2$
\affil{$^1$Nordita, KTH Royal Inst. of Technology and Stockholm University, Sweden}
\affil{$^2$Department of Engineering Sciences, Uppsala University, Sweden}
}

\begin{abstract}
Fractal grains have large surface area, which leads to more efficient condensation. The special limit case where the volume-area ratio is constant (corresponding to, e.g., a very rough grain surface or non-compacts aggregates) is particularly interesting, as well as convenient, from a mathematical point of view. If dust grains from AGB stars have `rough surfaces', it may have important implications for our understanding of dust and wind formation in AGB stars.
\end{abstract}
\section{Background and theory}
In dust condensation the accretion rate is proportional to the surface area of the grain, i.e., there is a difference between smooth, spherical and non-spherical grains. With rough fractal surfaces, the rate of accretion is obviously larger. So, could fractal grains grow fast enough to explain, e.g., recent results on supernova dust \citep[e.g.][] {Matsuura11, Gall14}? And then, how is dust formation in AGB stars affected?

The mass-growth rate of a dust particle of mass $m$ is ${dm/ dt} = \alpha_{\rm s} \langle v \rangle\, A_{\rm gr}\,m_{\rm X}\,n_{\rm X}(t)$
where $\alpha_{\rm s}$ is sticking coefficient (probability), $ \langle v \rangle$ is the thermal mean velocity of the growth species molecule X and $m_{\rm X}$,  $n_{\rm X}$ are its mass and number density, respectively.

\begin{figure}
  \resizebox{\hsize}{!}{
\includegraphics{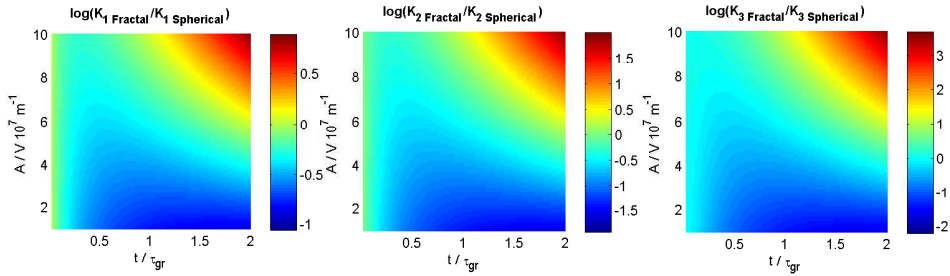}
}
\caption{\label{ratios} Comparison of solutions for smooth spherical grains ($h = 1$, $d = 2$) and fractal grains of extreme surface roughness ($h=0$, $d = 3$) for $\ell = 1,2,3$ and $\varsigma / 10^7\,{\rm m} \in [1,10]$. Note that $K_{3, \rm frac}$ (total dust mass) can be orders of magnitude larger or smaller than $K_{3, \rm sph}$ after two growth timescales $(t=2\, \tau_{\rm gr})$ depending on $\varsigma$.}
\end{figure}

Smooth spherical grains have a grain-surface area $A_{\rm gr} = 4\pi\,a^2 \propto m^{2/3}$, where $a$ is the radius. Assuming $dn_{\rm X}/dt \approx 0$ (reasonable before depletion of X) we find $m \propto t^3$. For a constant area-volume relation, i.e., $A_{\rm gr} = \varsigma\,V_{\rm gr} \propto m$, we find instead $m \propto e^t$. These are the two limit cases. A smooth sphere corresponds to a grain with a Hurst exponent $h=1$ and the latter case corresponds to $h=0$, where $h$ and the fractal dimension $d$ are related as $d = 1 + n - h$ for a self-similar surface in $n$-dimensional space \citep{Gneiting04}. Thus, for $n=2$ and $h \in [0,1]$ we have $d\in [2,3]$, which for (surface fractals) indicate a range of roughnesses from perfectly smooth to extremely rough and uneven. Real grains are likely somewhere in between the extreme cases $h = 0$ and $h = 1$, though. The equations for the moments of the grain-mass distribution $f(m,t)$ in case of smooth spherical grains ($h = 1$, $d = 2$) is then \citep{Gail88},
\begin{equation}
\label{eqnMoment1}
{dK_\ell\over dt} = {\ell}\,\xi(t)\,K_{\ell-1}(t), \quad K_\ell \equiv \int_{0}^{\infty} m^{\ell/3} f(m,t)\,dm, \quad n = 1,2,3,\dots,
\end{equation}
where $\ell$ is the moment order, $\xi \propto da/dt$ and $K_0 = n_{\rm d}$ is the dumber density of grains.
For the opposite limit ($h = 0$, $d = 3$) we derive the equations (the moments $K_\ell$ are still defined as above),
\begin{equation}
\label{eqnMoment2}
{dK_\ell\over dt} = {\ell}\,\tilde{\xi}(t)\,K_{\ell}(t),
\end{equation}
where $\tilde{\xi} \propto da/dt$, but with a different scaling factor compared to $\xi$ which includes $\varsigma$. Both sets of equations can be solved analytically up to the integral $\int_0^t n_{\rm X}(t')\,dt'$. In Fig. \ref{ratios} we compare the solutions for the spherical (Eq. \ref{eqnMoment1}) and the `fractal' case  (Eq. \ref{eqnMoment2}) for a range of $\varsigma$ values. The initial grain-mass distribution is a power-law with a slope $d\ln n/d\ln m = -0.5 \leftrightarrow d\ln n/d\ln a = -3.5$.

\section{Results and conclusions}
The moments (in particular the first and second) may initially increase slower in the `fractal limit' case compared to the case of smooth spherical grains. This effect depends on the adopted (constant) area-volume relation $\varsigma$, which is a number of the order $\sim\langle a\rangle^{-1}$, where $\langle a\rangle$ is the average grain radius. Regardless of the value of $\varsigma$ the {\it exponential} growth in the `fractal limit' ($h = 0$, $d = 3$) will eventually be faster than in any other case. For the range of $\varsigma$ considered here, `fractal' grains win the `mass-growth battle' within a growth timescale [$t \le \tau_{\rm gr} = \langle a(0)\rangle \,(da/dt)^{-1}$] when $\varsigma \gtrsim 5\cdot 10^7\,{\rm m}^{-1}$ and can reach a total dust mass that is up to $\sim 20$ times larger than for spherical grains. 

In summary: smooth spherical grains and fractal grains with constant $\varsigma$ represents two distinctly different regimes of dust condensation; dust grains in AGB star atmospheres with rough fractal surfaces (aggregates are unlikely), may condense significantly more efficient than in models of dust and wind formation, where grains are assumed to be smooth spheres.

\acknowledgements We thank Gunnar Niklasson for sharing his knowledge about fractal grains. 

\end{document}